\documentclass[twocolumn,showpacs,amsmath,amssymb,superscriptaddress,pra]{revtex4}

\usepackage{graphicx}

\newcommand{\beq}{\begin{equation}}
\newcommand{\eeq}{\end{equation}}
\newcommand{\beqa}{\begin{eqnarray}}
\newcommand{\eeqa}{\end{eqnarray}}

\newcommand{\ket}[1]{\left| #1 \right\rangle}
\newcommand{\bra}[1]{\left\langle #1 \right|}

\begin{document}

\title{Overcoming non-Markovian dephasing in single photon sources through post-selection}
\author{A. Nazir} \email{ahsan.nazir@ucl.ac.uk}
\affiliation{Department of Physics and Astronomy, University College London, Gower Street, London WC1E 6BT, U.K.}
\affiliation{Centre for Quantum Dynamics, Griffith University, Brisbane, Queensland 4111,
Australia}
\author{S. D. Barrett} \email{seandbarrett@gmail.com}
\affiliation{Blackett Laboratory, Imperial College London, Prince
Consort Road, London SW7 2BW, U.K.}

\date{\today}

\begin{abstract}
We study the effects of realistic dephasing environments on a pair of
solid-state single-photon sources in the context of 
the Hong-Ou-Mandel dip. By means of solutions for the Markovian or exact non-Markovian 
dephasing dynamics of the sources, we show that
the resulting loss of visibility depends crucially on the timing of
photon detection events. Our results demonstrate that the effective
visibility can be improved via temporal post-selection, and also that
time-resolved interference can be a useful probe of the interaction
between the emitter and its host environment.
\end{abstract}


\maketitle

Indistinguishable photons lie at the heart of linear optical~\cite{Kok07}  and distributed light-matter quantum information processing (QIP)~\cite{BarrettKok2005}. In particular, interference effects such as photon bunching and `which-path' erasure are central to a range of QIP protocols, leading to a recent drive for highly efficient single photon sources. Solid-state systems such as quantum dots (QDs), individual molecules, or colour centres in diamond~\cite{Shields2007} are emerging at the forefront of this new technology due to the potential they offer for realizing not only efficient but also controllable, deterministic single photon devices.  

However, solid-state systems are inherently noisy and thus their optical transitions are typically subject to energy fluctuations, for example due to coupling to phonon modes in the host system. These fluctuations tend to reduce the phase coherence of the emitted light, and hence degrade the visibility of interference effects~\cite{Bylander2003}. Dephasing can be viewed as broadening of the optical transition, which reduces the indistinguishability of the emitted photons. In the case of  QIP schemes~\cite{Kok07,BarrettKok2005}, this dephasing leads to logical errors in the calculation.  Overcoming this effect is important in view of the practical applications of single photon sources.  The realization of Fourier transform (i.e. lifetime) limited single photon devices has therefore become the goal for an extremely active area of research~\cite{Shields2007}. Moreover, understanding the interaction between the emitter and its host environment is an interesting problem in its own right.

\begin{figure}[!t]
\centering
\includegraphics[width=2.2in,height=1.5in]{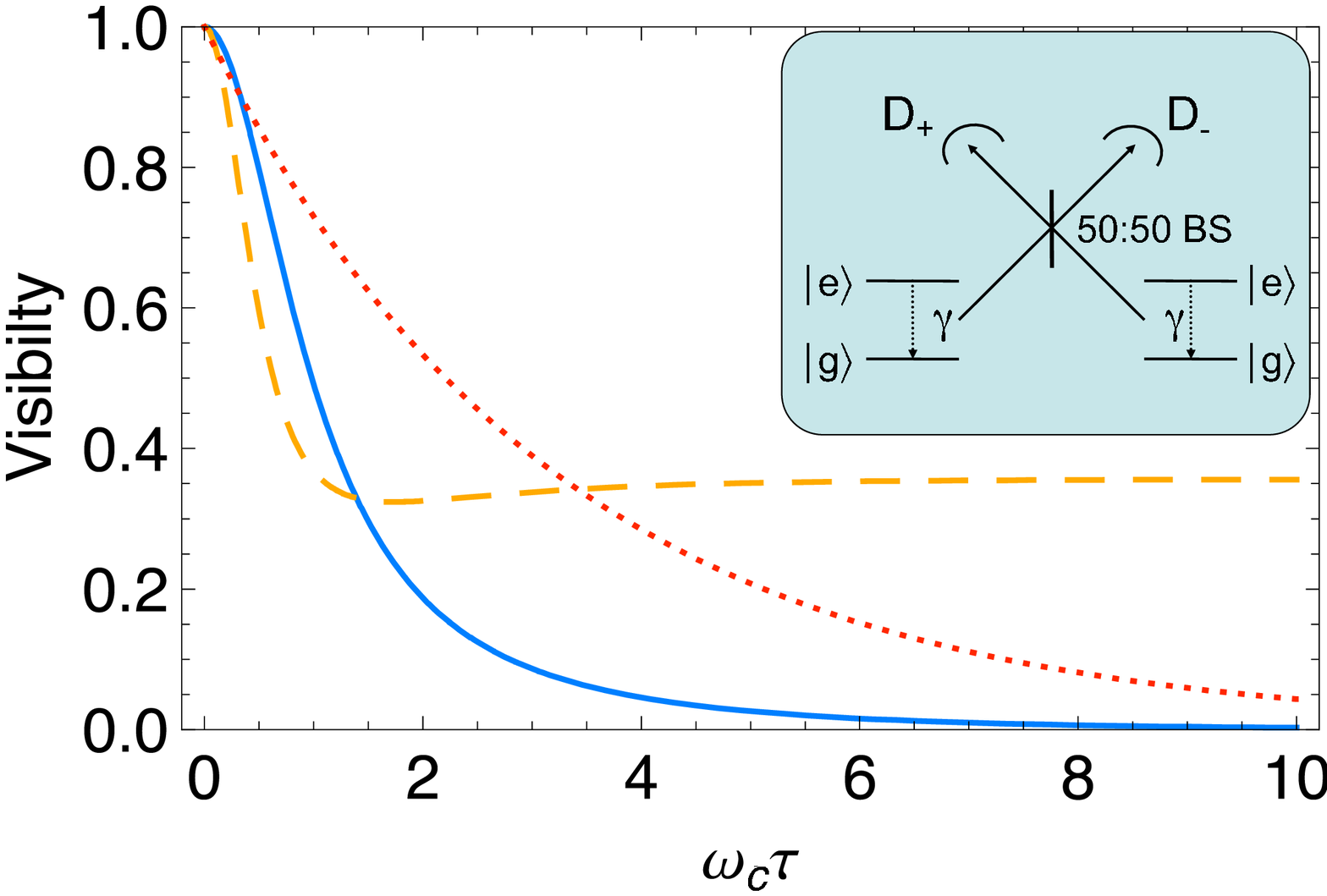}
\caption{Main: Visibility against $\omega_c\tau$  
for: an ohmic bath with $\Gamma_{n=1}(\tau)=A\ln{(1+\omega_c^2\tau^2)}+A\ln{(\sinh(\pi\tau/\beta)/(\pi\tau/\beta))}$  (blue, solid line); a superohmic bath with $\Gamma_{n=3}(\tau)=A\omega_c^2\tau^2(3+\omega_c^2\tau^2)/(1+\omega_c^2\tau^2)^2+A(\pi^2/3(\omega_c\beta)^2-1/(\omega_c\tau)^2+\pi^2{\rm csch}^2(\pi\tau/\beta)/(\omega_c\beta)^2)$  (yellow, dashed line); and Markovian dephasing of rate $\Gamma_{M}(\tau)=A\pi\tau/\beta$ (red, dotted line). Parameters: 
$A=0.5$, $\omega_c\beta=10$. Inset: Schematic of the interference setup. Two photons incident on a $50$:$50$ beam splitter (BS) interfere such that they are always detected at the same output, provided the photons themselves and the paths they take are identical. The resulting absence of coincidence counts {\it between} the detectors is known as the HOM dip. Its visibility measures the indistinguishability of the incident photons.}
\label{HOMvisibility}
\end{figure}

In this Rapid Communication we propose that these issues can be addressed with the aid of \emph{temporal post-selection} of the photons emitted from a noisy source. If sufficiently fast photodetectors are available, it is possible to post-select a sub-ensemble of detection events based on their detection times. By considering a  Hong-Ou-Mandel (HOM) type two-photon interference setup (see inset of Fig.~\ref{HOMvisibility}), we show that such post-selection can substantially improve the effective visibility in an interference experiment, albeit at the expense of effective source efficiency. This allows significant improvement in the fidelity of QIP protocols, at the expense of success probability, which nevertheless will be of benefit in near-future experiments with noisy sources. Furthermore, we show that post-selection can yield important information about the basic physics of the interaction between the emitter and its host environment. By measuring the HOM visibility as a function of time, one can directly probe the dynamics of the system-environment interaction.

We make use of an explicit model of a single photon source which is coupled both to the modes of the optical field, and to a solid-state environment. While the optical modes can be observed via single photon detectors, the experimentalist has no direct access to the environment degrees of freedom. This is a familiar situation in quantum optics, which could be dealt with using the standard technique of quantum trajectories~\cite{Plenio98}, whereby both the system-photon and system-environment couplings can be treated pertubatively within the Born-Markov (BM) approximation~\cite{Bylander2003,Plenio98,Stace03,Kiraz04}. However, although the BM approximation is typically very well satisfied for the system-photon coupling, in general, this approximation may break down for the system-environment coupling, and non-Markovian effects can be significant~\cite{Borri07}. Here, we move beyond the BM approximation for the solid-state environment, and find an \emph{exact analytic solution} for the dynamics induced by the system-environment coupling. We treat the system-photon coupling within the BM approximation, which allows us  to compute the observed photon statistics in a straightforward manner.

We model our single photon sources as a pair of two-level systems each with ground state $|g\rangle$, excited state $|e\rangle$, and equal energy splitting $E$. Interaction with the radiation field induces spontaneous decay from state $|e\rangle$ to $|g\rangle$ with associated photon emission.
In addition, both systems couple to their own bath of harmonic oscillators, representing the solid-state environment. We assume the sources are placed far enough apart that no interactions occur between them, either direct or mediated by the solid-state baths. A relevant physical implementation could, for example, be two optically active self-assembled QDs~\cite{Santori2002} or nitrogen-vacancy (NV) centres~\cite{Batalov08}, each coupled to their own cavity and operating within the Purcell regime~\cite{Kiraz04,Metz08}. Emitted photons are interfered on a $50$:$50$ beam splitter, without any delay introduced between the arrival times, and subsequent detection occurs at the photon counters $D_+$ and $D_-$. We will not treat imperfections in the setup such as limited photon detection and collection efficiency, or system frequency mismatch, in order to concentrate on the impact of the dephasing baths. However, in principle, such effects could be incorporated into the formalism~\cite{Metz08}. In fact, the only difference due to realistic detection efficiency would be a reduction in the total number of two-photon events; if one were to post-select only the sub-ensemble of events where two photons were observed from each excitation, one would obtain exactly the results presented here.

The Hamiltonian describing our setup is given by
$H=H_{0}+H_{\rm I}$, with $H_0=\sum_{j} H_{j,0}=\sum_j(E\ket{e}_j\bra{e}+H_{j,{\rm B}}+H_{j,{\rm X}})$ representing each two-level system and the harmonic oscillator baths in isolation ($j=1,2$ denotes the two qubits). Here, $H_{j,{\rm B}}=\sum_{{\bf k}}\omega_{j,{\bf k}}a_{j,{\bf k}}^{\dagger}a_{j,{\bf k}}$ and $H_{j,{\rm X}}=\sum_{{\bf q}}\nu_{j,{\bf q}}b_{j,{\bf q}}^{\dagger}b_{j,{\bf q}}$
describe the solid-state and photon environments, with frequencies $\omega_{j,{\bf k}}$ and $\nu_{j,{\bf q}}$ respectively ($\hbar=1$ throughout). The interaction term is given by $H_I=\sum_{j}H_{j,{\rm I}}=\sum_j(H_{j,{\rm SB}}+H_{j,{\rm SX}})$, where $H_{j,{\rm SB}}=\ket{e}_j\bra{e}\sum_{{\bf k}}(g_{j,{\bf k}}a_{j,{\bf k}}^{\dagger}+g_{j,{\bf k}}^*a_{j,{\bf k}})$ and $H_{j,{\rm SX}}=\sum_{{\bf q}}(\ket{g}_j\bra{e}f_{j,{\bf q}}b_{j,{\bf q}}^{\dagger}+\ket{e}_j\bra{g}f_{j,{\bf q}}^*b_{j,{\bf q}})$ account, respectively, for system-solid-state bath interactions via the excited states $\ket{e}_j$ with coupling $g_{j,{\bf k}}$, and for system-photon interactions (within the dipole and rotating-wave approximations~\cite{Carmichael93}) through the raising and lowering operators $\ket{e}_j\bra{g}$ and $\ket{g}_j\bra{e}$, with coupling $f_{j,{\bf q}}$. The bath creation (annihilation) operators $a_{j,{\bf k}}^{\dagger}$ ($a_{j,{\bf k}}$) and $b_{j,{\bf q}}^{\dagger}$ ($b_{j,{\bf q}}$) obey boson statistics, and none of the baths share any common modes. We note that the form of system-solid-state bath interaction ($H_{j,\rm{SB}}$) assumed above is particularly relevant to the study of short-time dephasing behaviour~\cite{Solenov07}, such as that observed experimentally due to phonons in QDs~\cite{Borri07}.

Moving into the interaction picture with respect to $H_{0}$, the exact evolution of the combined system-environment density operator $\chi_{\rm I}(t)=e^{iH_0t}\chi(t)e^{-iH_0t}$ is given by~\cite{Carmichael93}
\begin{equation}\label{combinedevolve}
\dot{\chi}_{\rm I}=-i[H_{\rm I}(t),\chi(0)]-\int_0^t d\tau[H_{\rm I}(t),[H_{\rm I}(\tau),\chi_{\rm I}(\tau)]],
\end{equation}
where $H_{\rm I}(t)=e^{iH_0t}H_{\rm I}e^{-iH_0t}=\sum_j(H_{j,{\rm SB}}(t)+H_{j,{\rm SX}}(t))$. We are interested here in the regime in which the system-solid-state bath dynamics is slow on the timescale set by the radiation field correlation time (which is typically extremely short, $\tau_c\sim1/E\sim1$~fs for $E=1$~eV~\cite{Hohenester04}). In this case, we may perform a Born approximation on the density operator factorizing it into a system-solid-state bath contribution $\rho$ and a radiation field part $R$ as $\chi=\rho\otimes R_1\otimes R_2$, assuming the baths are initially uncorrelated. It is important to stress that we do not further factorize $\rho$ and so fully account for correlations built up between the system and solid-state bath during their combined evolution. Inserting this form for 
$\chi$ into Eq.~(\ref{combinedevolve}) and tracing over the photon baths we find the simpler form
$\dot{\rho}_{\rm I}=-i\sum_j[H_{j,\rm SB}(t),\rho_{\rm I}(t)]-\sum_j\int_0^t d\tau{\rm Tr}_{j,\rm{X}}[H_{j,{\rm SX}}(t),[H_{j,{\rm SX}}(\tau),\rho_{\rm I}(\tau)R_j]]$,
where we assume that $R_j$ are thermal equilibrium states. The interpretation of this equation is clear; the first term describes the exact evolution of $\rho_{\rm I}$ due solely to $H_{1,{\rm SB}}$ and $H_{2,{\rm SB}}$, while, within a Markov approximation, the second term gives rise to the usual Lindblad operators associated with spontaneous emission and defines the transition rates $\gamma_{j}$ for each qubit~\cite{Plenio98,Carmichael93}.

The quantum jump formalism is therefore particularly well suited to studying the dynamics of our system, here subject to continual observation by detectors $D_{\pm}$. We define jump operators corresponding to photon emission and detection as $c_j=\sqrt{\gamma_j}\ket{g}_j\bra{e}$, projecting each two-level system from its excited to ground-state (for simplicity, 
we take $\gamma_1=\gamma_2=\gamma$). The action of the beam splitter is to mix these operators, erasing the information on where the emission originated, and leading to $c_{\pm}=(c_1\pm c_2)/\sqrt{2}=\sqrt{\gamma/2}\left(\ket{g}_1\bra{e}\pm\ket{g}_2\bra{e}\right)$,
as jump operators for a photon count in $D_{\pm}$, respectively. Assuming the initial state of the sources to be $\ket{e}_1\otimes\ket{e}_2=\ket{ee}$, and that all emitted photons are collected and detected, within any increment of time $dt$ the system may evolve in one of only two possible ways. If no photon is observed, we know that neither source has returned to its ground-state. In this case, the combined system-solid-state bath evolution is generated by the conditional no-jump master equation $\dot{\rho}_{\rm I}=-i(H_{\rm eff}(t)\rho_{\rm I}-\rho_{\rm I} H_{\rm eff}^{\dagger}(t))$, with non-Hermitian effective Hamiltonian
\begin{equation}\label{Heff}
H_{\rm eff}(t)=\sum_{j}H_{j,{\rm SB}}(t)-(i/2)c_+^\dagger c_+-(i/2)c_-^\dagger c_-.
\end{equation} 
We define the corresponding non-unitary, no-jump time-evolution operator by
\begin{equation}\label{nojumpevolve}
U_{\rm I}(t_i,t_f)={\cal T}\exp{\left(-i\int_{t_i}^{t_f}dtH_{\rm eff}(t)\right)},
\end{equation}
where ${\cal T}$ is the time-ordering operator. The state generated by $\tilde{\rho}_{\rm I}(t_f)=U_{\rm I}(t_i,t_f)\rho_{\rm I}(t_i)U_{\rm I}^{\dagger}(t_i,t_f)$ will be unnormalised and ${\rm Tr}(\tilde{\rho}_{\rm I}(t_f))$ can thus be interpreted as the probability that no detector clicks are observed within the time interval $t_i$ to $t_f$. Accounting for the effect of the time-ordering operator in $U_{\rm I}(t_i,t_f)$ is straightforward if we employ a Magnus expansion of the exponential~\cite{Magnus54}, since all terms of the form $[H_{\rm eff}(t),[H_{\rm eff}(t'),H_{\rm eff}(t'')]]$ and higher disappear, allowing us to solve for the exact dynamics due to the solid-state environment~\cite{Mahan00}.

A click in either detector, on the other hand, 
implies a discontinuous 
lowering of the system state and we apply the appropriate jump operator
\begin{equation}\label{jumpevolve}
\rho_{\rm I}(t+dt)=c_{\pm}\rho_{\rm I}(t)c_{\pm}^{\dagger}/{\rm Tr}(c_{\pm}^{\dagger}c_{\pm}\rho_{\rm I}(t)),
\end{equation}
for a count in $D_{\pm}$. Here, ${\rm Tr}(c_{\pm}^{\dagger}c_{\pm}\rho_{\rm I}(t))$ is the probability that a click occurs in detector $D_{\pm}$ in the interval $t$ to $t+dt$ given that that the system is in a normalized state $\rho_{\rm I}(t)$ at time $t$.

Eqs.~(\ref{Heff}-\ref{jumpevolve}) allow us to calculate the full time-resolved statistics of the photodetector counts, and thereby assess the impact of source dephasing on the visibility of the HOM dip. The sources are again assumed to be 
initialised in state $\ket{ee}$, with the baths in thermal equilibrium, $\rho_{j,{\rm B}}=e^{-\beta H_{j,{\rm B}}}/{\rm Tr}(e^{-\beta H_{j,{\rm B}}})$, where $\beta=1/k_BT$,  $k_B$ being the Boltzmann constant. The combined system-bath density operator is thus of initial form $\rho(0)=\ket{ee}\bra{ee}\otimes\rho_{1,{\rm B}}\otimes\rho_{2,{\rm B}}$. It is instructive to first calculate the intermediate state of the system, $\tilde{\rho}_S$, at time $t_2$, conditional on observing the first detector click at time $t_1 \le t_2$ in detector $D_+$, but before the second click. From Eqs. (\ref{Heff}-\ref{jumpevolve}), on tracing over both environments, we find
\begin{align}\label{rhot2primed}
\tilde{\rho}_S&=\frac{e^{-\gamma\tau}}{2}\Big[\ket{ge}\bra{ge}+\ket{eg}\bra{eg}+e^{-(\Gamma_1(\tau)+\Gamma_2(\tau))}\nonumber\\
&\quad\times(e^{i\phi(t_1,t_2)}\ket{ge}\bra{eg}+e^{-i\phi(t_1,t_2)}\ket{eg}\bra{ge})\Big],
\end{align}
where $\tau=t_2-t_1$ is the time elapsed since the first click. Here, $\Gamma_j(\tau)=\sum_{\bf k}|\alpha_{j,{\bf k}}(t_1,\tau)/\sqrt{2}|^2\coth{(\beta\omega_{j,{\bf k}}/2)}$, with $\alpha_{j,{\bf k}}(t_1,\tau)=(g_{j,{\bf k}}/\omega_{j,{\bf k}})e^{i\omega_{j,{\bf k}} t_1}(1-e^{i\omega_{j,{\bf k}}\tau})$,  are time-dependent decoherence functions, characterising the undesirable effects of the baths on the system evolution. The phase factor is $\phi(t_1,t_2)=\Lambda_2(t_1,t_2)-\Lambda_1(t_1,t_2)$, where $\Lambda_j(t_1,t_2)=\sum_{{\bf k}}|g_{j,{\bf k}}/\omega_{j,{\bf k}}|^2(\omega_{j,{\bf k}}\tau+2\sin{\omega_{j,{\bf k}}t_1}-2\sin{\omega_{j,{\bf k}}t_2}+\sin{\omega_{j,{\bf k}}\tau})$. We restrict ourselves to the case when both sources couple equally to the same form of environment, giving $\Gamma_1(\tau)=\Gamma_2(\tau)=\Gamma(\tau)$ and $\phi(t_1,t_2)=0$. This is realized experimentally when two consecutive photons are interfered from the same source~\cite{Santori2002}. 

The probability density of the second click being observed in either the same detector $D_+$ ($=p(t_2,+|t_1,+)$) or opposite $D_-$ ($=p(t_2,-|t_1,+)$) during a time interval $t_2$ to $t_2+dt$, conditional on the first click occurring in $D_+$ at time $t_1$, may now be calculated from $p(t_2,\pm|t_1,+)={\rm Tr}(c_{\pm}^{\dagger}c_{\pm}\tilde{\rho}_S(\tau))$. We find
\begin{equation}\label{HOMP+-}
p(t_2,\pm|t_1,+)=p(\pm|+,\tau)=\frac{\gamma}{2}e^{-\gamma\tau}\left[1\pm e^{-2\Gamma(\tau)}\right],
\end{equation}
which reduces to $p(+|+,\tau)=\gamma e^{-\gamma\tau}$ and $p(-|+,\tau)=0$ in the absence of dephasing, then implying that both photons exit at the same port of the beamsplitter (i.e. perfect HOM interference). 
Dephasing of the sources therefore destroys the interference effect, and by implication the indistinguishability of the emitted photons, by an amount that depends not only on the form and strength of the system-bath interactions, but also on the {\it difference in detection times} $\tau$. This may be seen more clearly by considering the visibility of the HOM dip, which we define in the ideal detector limit to be
\begin{align}\label{visibility}
\nu(\tau)=\left|\frac{p(+|+,\tau)-p(-|+,\tau)}{p(+|+,\tau)+p(-|+,\tau)}\right|
=e^{-2\Gamma(\tau)},
\end{align}
measuring the normalised difference in conditional detection probabilities for $D_+$ and $D_-$ at time $t_2$. In the absence of any dephasing $\nu(\tau)=1$ for all $\tau$ as we consider no further imperfections in the setup, while in general $\nu(\tau)$ is completely determined by the decoherence function $\Gamma(\tau)$. There is no dependence on the spontaneous emission rate at the perfectly time-resolved level. 

We plot the visibility as a function of detector click separation 
in Fig.~\ref{HOMvisibility}. To illustrate the general behaviour, we cast the decoherence function in terms of the spectral density~\cite{Leggett87}, $J(\omega)=\sum_{{\bf k}}|g_{{\bf k}}|^2\delta(\omega-\omega_{{\bf k}})=A\omega^n e^{-\omega/\omega_c}$, as $\Gamma(\tau)=\int_0^{\infty}(J(\omega)/\omega^2)(1-\cos{\omega\tau})\coth{(\beta\omega/2)}$. Here, $\omega_c$ is a high frequency cut-off and $A$ an appropriate coupling strength. We consider both ohmic ($n=1$) and superohmic ($n=3$) spectral densities as they are relevant, for example, to carrier-phonon interactions in the solid-state~\cite{Mahan00}. We also plot the performance with a Markovian rate for comparison, where $\Gamma_M(\tau)$ is given by the thermal limit of the ohmic case. We see that for both the ohmic and Markovian cases the visibility tends asymptotically to zero as the click separation becomes large, although the ohmic behaviour differs qualitatively on shorter timescales due to non-Markovian effects. In the case of a superohmic spectral density the evolution is entirely non-Markovian, with remnant visibility $\nu_{n=3}(\infty)=\exp{\left(-2 A(1+\pi^2/3(\omega_c\beta)^2)\right)}$.

\begin{figure}[!t]
\centering
\includegraphics[width=2.0in,height=1.2in]{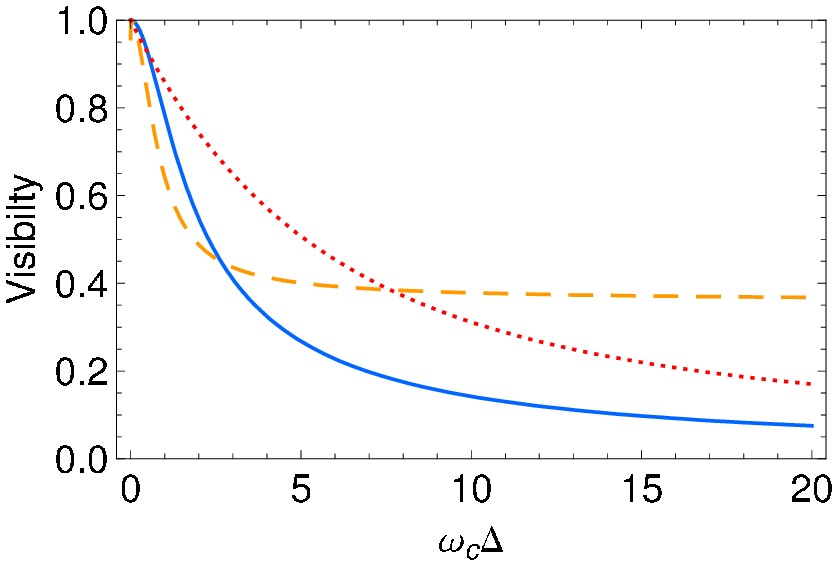}
\caption{Visibility against $\omega_c\Delta$ for the ohmic (blue, solid line), superohmic (yellow, dashed line), and Markovian (red, dotted line) baths. Parameters are as in Fig.~\ref{HOMvisibility}, with $\gamma/\omega_c=0.01$.}
\label{testfig2}
\end{figure}

Clearly, for ideal detectors of time resolution much shorter than any system-bath dynamics, Fig.~\ref{HOMvisibility} suggests that by post-selecting detection events closely spaced in time~\cite{Legero04} the visibility of the two-photon interference may be improved, overcoming noise inherent within the sources. In fact, this is also true for non-identical sources, where the visibility becomes $\nu=e^{-(\Gamma_1(\tau)+\Gamma_2(\tau))}|\cos{\phi(t_1,t_1+\tau)}|$. Here, post-selection will still improve the quality of the interference, though now it is not simply $\tau$ that determines the loss in visibility, but also the detection time $t_1$ itself. Although the new factor in the visibility is oscillatory, in practice it is impossible to measure the detection time with arbitrarily high precision. Hence, we propose setting a detection time post-selection window. In this situation, for non-identical sources, it is generally advantageous not only to post-select closely spaced detection events, but those in which the detection time $t_1$ is short as well. 

Returning to the case of identical sources, we illustrate this idea in Fig.~\ref{testfig2}, where a modified visibility function $\nu'(\Delta)=|(p_{++}(\Delta)-p_{+-}(\Delta))/(p_{++}(\Delta)+p_{+-}(\Delta))|$ is plotted. Here, $p_{++}(\Delta)=\int_0^{\omega_c\Delta}d\tau p(+|+,\tau)$ and $p_{+-}(\Delta)=\int_0^{\omega_c\Delta}d\tau p(-|+,\tau)$ are the relevant conditional probability densities now integrated over the window width $\omega_c\Delta$.
We see that, in principle, the improvement due to post-selection can be extremely good since the visibility will always approach unity as $\Delta\rightarrow0$.  
However, any improvement necessarily comes at the cost of effective source efficiency and practical considerations on the number of photons that can reasonably be discarded limits what would be possible for any particular setup. It is important to stress that such an improvement can only be observed in a time-resolved experiment~\cite{Legero04}, though even if this resolution is fairly poor some improvement should be possible when there is a long-time decay of visibility towards a fixed value, as in the ohmic and Markovian examples considered 
here. In contrast, in the `bad detector' limit of zero resolution, we must integrate the conditional probability densities over all $\tau$. Here, assuming a set system-bath interaction strength, the simplest way to improve the visibility is to engineer a larger effective decay rate such that the system emits on timescales shorter than those of the dominant dephasing processes~\cite{Varoutsis05}.

In the opposite limit, when the system spontaneous emission rate is small compared to the typical rate for dephasing~\cite{Borri07}, the form of the system-bath interaction can itself be probed with fast detectors by successively widening the post-selection window (scanning the visibility in time as in Fig~\ref{testfig2}). Provided that the post-selection window size is kept much shorter than the excited state lifetime ($\gamma\Delta\ll1$),  then the visibility has essentially no dependence on the value of $\gamma$ itself, allowing the bath-induced dynamics to be isolated and characterised independently from spontaneous decay processes. For example, in this regime the visibility is well described by $\nu'_{M}(\Delta)=\beta(1-e^{-2A\pi\Delta/\beta})/2A\pi\Delta$ in the Markovian case, and by $\nu'_{n=1}(\Delta)=B_{-\Delta^2}(1/2,1-2A)/2i\Delta$ in the low temperature ohmic case, where $B_z(a,b)$ is the incomplete Beta function. Such measurements therefore offer the potential for clear signatures of deviations from Markovian dephasing behaviour, as well as a means to assess system-bath coupling strengths.

To summarise, we have shown analytically how to incorporate both Markovian and non-Markovian dephasing baths into the quantum jump formalism for emitting two-level systems. We have analysed the impact of such baths on HOM interference from a pair of solid-state single photon sources, demonstrating that the dephasing may be overcome by post-selection providing that sufficient time-resolution exists within the photodetectors. In general, time-resolved two-photon interference can provide significant insight into solid-state decoherence processes. As fast detectors down to sub-picosecond time resolution are now available~\cite{Kuzucu08}, much shorter than the typical excited state lifetimes of nanoseconds or greater in, for example, QDs or NV centres, experiments to explore such phenomena are feasible with current technology in a number of different physical systems, and will become more so as technology improves. Furthermore, it should also be possible to apply our formalism to gauge the impact of dephasing on a wide range of distributed entangling protocols that exploit photon interference effects.

We thank H. Wiseman, T. Stace, J. Rarity, J. O'Brien, and G. Rempe for stimulating discussions. AN is supported by a Griffith University Postdoctoral Fellowship, the State of Queensland, the Australian Research Council, and by the {\sc epsrc}.  SDB is supported by the {\sc
epsrc}.

\end{document}